\documentclass[10pt,showpacs]{iopart}

\usepackage{iopams}  
\usepackage{graphicx}
\usepackage{bm}
\expandafter\let\csname equation*\endcsname\relax
\expandafter\let\csname endequation*\endcsname\relax

\usepackage{amsmath,multicol,cuted}
\usepackage[font=small]{caption}
\nopagebreak

\begin{document}
\title[Current fluctuations across a nano-pore]{Current fluctuations across a nano-pore}

\author{Mira Zorkot and Ramin Golestanian}
\address{Rudolf Peierls Centre for Theoretical Physics, Oxford University, Oxford, OX13NP, United Kingdom}
\ead{ramin.golestanian@physics.ox.ac.uk}
\vspace{10pt}
\begin{indented}
\item[]November 2017
\end{indented}
\renewcommand{\rmd}{\textrm{d}}
\newcommand{\kT}[0]{k_{\textrm{B}}T}
\newcommand{\myvec}[1]{\boldsymbol{#1}}
\newcommand{\myten}[1]{\mathsf{#1}}
\begin{abstract}
The frequency-dependent spectrum of current fluctuations through nano-scale channels is studied using analytical and computational techniques. Using a stochastic Nernst-Planck description and neglecting the interactions between the ions inside the channel, an expression is derived for the current fluctuations, assuming that the geometry of the channel can be incorporated through the lower limits for various wave-vector modes. Since the resulting expression turns out to be quite complex, a number of further approximations are discussed such that relatively simple expressions can be used for practical purposes. The analytical results are validated using Langevin dynamics simulations.
\end{abstract}
\vspace{2pc}
\submitto{\JPCM}
\maketitle
\ioptwocol
\maketitle

\section{Introduction}
With recent advances in the development of synthetic and biological nano-pore science and its potential for technological applications such as in DNA sequencing, understanding the fundamental properties of the dynamics of ionic transport is of paramount importance {\cite{kasianowicz1996,2000_Deamer,2000_Meller,Kong2016}}. It is widely established that the accuracy of the DNA and RNA sequencing methods that rely on transport in nano-pores is limited by the prevalent $1/f$ noise in the ionic current {\cite{laszlo2014decoding,1988_Weissman_RevModPhys,2015_Heerema}}, although there have been theoretical proposals to take advantage of the unique properties of the noise in these systems to speed up the process of sequencing by orders of magnitude \cite{Jack1,Jack2,Jack3} . Understanding the source of the noise in nano-pores has been a subject of debate for decades {\cite{2007_Tabard-Cossa,Goy2,Goy3,Siwy}}. There have been experiments conducted on biological nano-pores, investigating whether the movements of the subunits of the channel walls can be identified as the source of the noise {\cite{Siwy,Dekker,Neher,bezrukov1993,2000_Bezrukov_PRL,1997_Wohnsland}}. Experiments performed on relatively flexible solid state nano-pores could not conclusively attribute the magnitude of the noise to the deformation fluctuations, the surface charge on the pore wall, or the geometry {\cite{tasserit,li2001,smeets2006,chen2004,2009_Hoogerheide,2009_Powell-Siwy_PRL}}. Earlier suggestions that the noise is dominated by number fluctuations as observed in metal samples {\cite {1970_Hooge}} did not appear to be consistent with observations in solid state and Kapton graphene nano-pores {\cite{tasserit,Dekker,smeets2006,2008_Smeets_PNAS}}. Despite various propositions based on different characteristics of the pore, and a limited number of theoretical proposals  {\cite{2008_Fulinski}}, a consistent picture that can explain all the observed features is still unidentified. 
 
\begin{figure}[t]
\centering
\includegraphics[width=1.0\columnwidth]{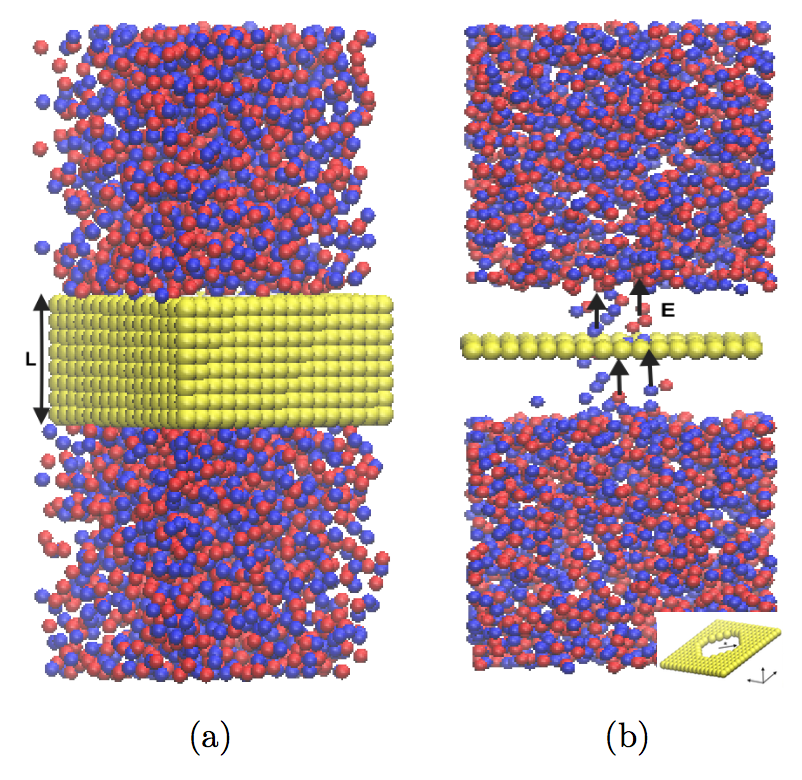}
\caption{(a) A simulation snapshot showing the impermeable membrane characterized by length $L$ and the ion reservoirs at the top and bottom of the membrane, with blue representing negative ions and red representing positive ions. (b) The slice across which charges are being counted and the current flow is measured. The electric field $E$ is applied on every ion crossing the pore. Inset: a cross section that shows the pore radius.}
\label{fig:setup}
\end{figure}

We have recently performed a theoretical analysis and demonstrated that ion interactions play a crucial role in the frequency dependence of the current fluctuations at small frequencies while at higher frequencies the dependence in controlled by other parameters such as geometric features, concentration of ions, the electric field and so on {\cite{zorkot2016-1,zorkot2016-2}}. The analytical treatment presented in Refs. {\cite{zorkot2016-1,zorkot2016-2}} were simplified such that a compact closed form expression can be used for comparison with simulation data. Here, we present the full analytical description of the system and show how using various approximations we can simplify it to the expressions used in the above references. The rest of the paper is organized as follows. In section \ref{sec:theory}, the analytical description of the stochastic dynamics of the ionic concentrations is presented and used to study ionic current fluctuations. Section \ref{sec:simulation} describes our Langevin dynamics simulations (see figure \ref{fig:setup} for typical snapshots) and the comparison between the theoretical results and the current fluctuations as measured from the simulations. Further possible approximation schemes and simplifications are examined in section \ref{sec:simple_theory}. 

\section{Theoretical Formulation of the Stochastic Dynamics}\label{sec:theory}

Monovalent ions are considered to flow through a cylinder of length $L$ and radius $R$; see figure \ref{fig:setup}. The positive and negative (henceforth abbreviated as $\pm$) ions undergo a stochastic dynamics described by trajectories ${\bm{r}}_i^\pm(t)$, under the influence of an applied electric field ${\bm{E}}$ (that will be taken to be along the pore) while they are inside the channel. The dynamics of the ions can be characterized by fluctuating concentrations $C^{\pm}({\bm{x}},t)=\sum_i \delta^3({\bm{x}}-{\bm{r}}_i^\pm(t))$ and the corresponding flux densities ${\bm{J}}^{\pm}({\bm{x}},t)$, where ${\bm{x}}$ denotes the position in three dimensions and $t$ denotes the time. We describe the flow of ions through the pore using the continuity equation
\begin{equation} \label{eqn:fick}
\frac{\partial C^{\pm}}{\partial t} + {\bm{\nabla}} \cdot \bm{J}^{\pm} = 0,
\end{equation}
where the fluctuating fluxes are defined as
\begin{equation}\label{eqn:fokk}
\myvec{J}^{\pm} = -D {\bm{\nabla}} C^{\pm} \pm \mu e C^{\pm}({\bm{E}}-{\bm{\nabla}}\phi)-{\bm{\eta}}^\pm.
\end{equation}
Here, ${\bm{\eta}}^{\pm}$ represent the thermal noise, $e$ denotes the elementary charge, $\mu$ is the mobility of ￼the ions (taken to be equal for both valences {\cite{weast1988}}), and the electrostatic potential $\phi$ satisfies the Poisson equation:
\begin{equation}\label{eq:poisson0}
-\nabla ^2 \phi=\frac{4\pi e}{\epsilon}(C^{+}-C^{-}),
\end{equation}
where $\epsilon$ is the dielectric constant of the medium. 

To simplify the notation, we introduce the dimensionless electrostatic potential $\psi=\frac{e\phi}{\kT}$, and write the Poisson equation as
\begin{equation}\label{eq:poisson}
-\nabla ^2 \psi=~4\pi \ell_{\rm B}(C^{+}-C^{-}),
\end{equation}
in terms of the Bjerrum length $\ell_{\rm B}\equiv \frac{e^2}{\epsilon \kT}$, which is about $7\,{\rm \AA}$ for water at room temperature. In this notation, the expressions for the fluxes simplify as follows
\begin{equation}\label{eqn:fokk-2}
\myvec{J}^{\pm} = -D {\bm{\nabla}} C^{\pm} \pm D C^{\pm}({\bm{\mathcal{E}}}-{\bm{\nabla}}\psi)-{\bm{\eta}}^\pm,
\end{equation}
where we define a reduced electric field ${\bm{\mathcal{E}}}=\frac{e{\bm{E}}}{\kT}$. \\

\subsection{Stochastic Fluctuations in Fourier Space}\label{sec:stoch-fourier}

We are interested in the fluctuations in the ionic concentrations around the average concentration $C_{\rm{pore}}$, defined via $C^{+}=C_{{\rm{pore}}}+\delta C^{+}$ and $C^{-}=C_{{\rm{pore}}}+\delta C^{-}$, where we expect the fluctuations to be relatively small, i.e. $\delta C^\pm \ll C_{\rm{pore}}$. Using the average concentration of the ions inside the pore, we define the Debye length $\kappa^{-1}$ through $\kappa^2=8\pi\ell_{\rm B}C_{{\rm{pore}}}$.

To derive an expression for the current density fluctuations, we start by taking the Fourier transform of equation (\ref{eq:poisson}):
\begin{equation}
\psi({\bm{q}},\omega)=\dfrac{4\pi\ell_{\rm B}}{q^2}\big[\delta C^{+}({\bm{q}},\omega)-\delta C^{-}({\bm{q}},\omega)\big],
\end{equation}
which we insert in the Fourier transform of equations (\ref{eqn:fick}) and (\ref{eqn:fokk-2}), to obtain a set of two coupled equations for the concentration fluctuations as follows:
\begin{strip}
\begin{eqnarray}
i\omega \delta C^{\pm}({\bm{q}},\omega) & = & - D q^2\delta C^{\pm}({\bm{q}},\omega)\pm i D \; ({\bm{q}}\cdot{\bm{\mathcal{E}}}) \;\delta C^{\pm}({\bm{q}},\omega)-i{\bm{q}}\cdot{\bm{\eta}}^{\pm}({\bm{q}},\omega)\nonumber \\
&\mp & 4\pi\ell_{\rm B}D\int\dfrac{\rmd \Omega}{2\pi}\dfrac{\rmd^3 {\bm{k}}}{(2\pi)^3}\Big[\dfrac{{\bm{q}}\cdot({\bm{q}}-{\bm{k}})}{({\bm{q}}-{\bm{k}})^2}\Big]\delta C^{\pm}({\bm{k}},\Omega)\big[\delta C^{+}({\bm{q}}-{\bm{k}},\omega-\Omega)-\delta C^{-}({\bm{q}}-{\bm{k}},\omega-\Omega)\big].\label{eq:cc}
\end{eqnarray}
\end{strip}
This process results in second order nonlinearities in our stochastic field equations that govern the dynamics of the ionic density fluctuations. The nonlinearities are the result of the Coulomb interactions between the mobile ions, and responsible for the $1/f$ behaviour observed in the current fluctuations at very small frequencies \cite{zorkot2016-1,zorkot2016-2}. The structure of the nonlinear stochastic field theory is reminiscent of the recently studied model for the homeostasis of a colony of cells with growth and chemical signalling, which has been studied using dynamical renormalization group (RG) methods \cite{gelimson}. We plan to perform a similar analysis with the current system in the future. However, for the purpose of the present investigation, we would like to focus on the linear theory only, and therefore, we will ignore the nonlinear terms in what follows. We thus expect the resulting theoretical scheme to correctly describe the high frequency part of the noise spectrum.

It is now helpful to define the charge density $\rho$, the net number density $c$, and their corresponding fluxes as follows:
\begin{eqnarray}
\hskip2cm&& \rho=\delta C^{+}-\delta C^{-}, \nonumber \\
&& c=\delta C^{+}+\delta C^{-}, \nonumber \\
&& {\bm{J}}_{\rho}={\bm{J}}^{+}-{\bm{J}}^{-},\nonumber \\
&& {\bm{J}}_{c}={\bm{J}}^{+}+{\bm{J}}^{-}.\nonumber
\end{eqnarray}
In particular, we are interested in the expression for the current density ${\bm{J}}_{\rho}(q,\omega)$ and the particle flux  ${\bm{J}}_{c}(q,\omega)$:
\begin{eqnarray}
\hskip0.5cm
{\bm{J}}_{\rho}({\bm{q}},\omega)&=& 2 D C_{{\rm{pore}}}{\bm{\mathcal{E}}}(2\pi)\delta(\omega)(2\pi)^3\delta^3({\bm{q}})  \nonumber \\
&&+i D {\bm{q}}\left(1+\frac{\kappa^2}{q^2}\right) \rho({\bm{q}},\omega)+D{\bm{\mathcal{E}}} c({\bm{q}},\omega) \nonumber \\
&&-\big[ {\bm{\eta ^{+}}}({\bm{q}},\omega) - {\bm{\eta ^{-}}}({\bm{q}},\omega) \big],\label{eqn:currentd}\\ \nonumber \\
\hskip0.5cm
{\bm{J}}_{c}({\bm{q}},\omega)&=& i D {\bm{q}} c({\bm{q}},\omega)+ D {\bm{\mathcal{E}}}\rho({\bm{q}},\omega) \nonumber \\
&&-\big[ {\bm{\eta ^{+}}}({\bm{q}},\omega) + {\bm{\eta ^{-}}}({\bm{q}},\omega) \big].\label{eqn:particle-flux-d}
\end{eqnarray}
Therefore, we can first solve for the two concentrations and insert them in the above expressions to obtain the fluxes.

Rearranging the expressions in equation (\ref{eq:cc}) without the nonlinear terms, and rewriting them in terms of $\rho$ and $c$, we obtain:
\begin{equation}
\begin{cases}
\left.\begin{aligned}
&(i\omega+D q^2) \; c({\bm{q}},\omega)-i D \; ({\bm{q}}\cdot{\bm{\mathcal{E}}}) \;\rho({\bm{q}},\omega)=\\
&\hskip3cm -i{\bm{q}}\cdot\Big[{\bm{\eta}}^+({\bm{q}},\omega)+{\bm{\eta}}^-({\bm{q}},\omega)\Big],\\
\end{aligned}\right.\\
\begin{aligned}
&[i\omega+D (q^2+\kappa^2)] \; \rho({\bm{q}},\omega)-i D \; ({\bm{q}}\cdot{\bm{\mathcal{E}}}) \; c({\bm{q}},\omega)=\\
& \hskip 3cm -i{\bm{q}}\cdot\Big[{\bm{\eta}}^+({\bm{q}},\omega)-{\bm{\eta}}^-({\bm{q}},\omega)\Big].\nonumber
\end{aligned}
\end{cases}
\end{equation}
We can then use matrix algebra to obtain the following results
\begin{eqnarray}
\hskip1cm
c({\bm{q}},\omega)&=&G^+_{c}({\bm{q}},\omega) \left[-i{\bm{q}} \cdot {\bm{\eta}}^{+}({\bm{q}},\omega)\right] \nonumber \\
&+&G^-_{c}({\bm{q}},\omega) \left[-i{\bm{q}} \cdot {\bm{\eta}}^{-}({\bm{q}},\omega)\right], \label{eqn:crho-c}\\ \nonumber \\
\hskip1cm
\rho({\bm{q}},\omega)&=&G^+_{\rho}({\bm{q}},\omega) \left[-i{\bm{q}} \cdot {\bm{\eta}}^{+}({\bm{q}},\omega)\right] \nonumber \\
&-&G^-_{\rho}({\bm{q}},\omega) \left[-i{\bm{q}} \cdot {\bm{\eta}}^{-}({\bm{q}},\omega)\right],\label{eqn:crho-rho}
\end{eqnarray}
where the response functions are defined as follows
\begin{equation}
\begin{cases}\nonumber
G^\pm_{c}({\bm{q}},\omega)={\cal M}({\bm{q}},\omega) \left[i\omega+D (q^2+\kappa^2) \pm iD{\bm{q\cdot\mathcal{E}}}\right], \\\\
G^\pm_{\rho}({\bm{q}},\omega)={\cal M}({\bm{q}},\omega) \left[i\omega+D q^2 \pm iD{\bm{q\cdot\mathcal{E}}}\right], \\
\end{cases}
\end{equation}
with 
\begin{equation}
{\cal M}^{-1}={[i\omega+Dq^2][i\omega+D(q^2+\kappa^2)]+D^2(\bm{q}\cdot{\bm{\mathcal{E}}})^2}.
\end{equation}
Inserting equations (\ref{eqn:crho-c}) and (\ref{eqn:crho-rho}) in equations (\ref{eqn:currentd}) and (\ref{eqn:particle-flux-d}), we find the following expressions for the flux densities, which we present in components using index notation:
\begin{eqnarray}
J_{\rho i}&=&\left[D\left(1+{\kappa^2}/{q^2}\right) q_i q_j G^+_{\rho}-i D {\cal E}_i q_j G^+_{c}-\delta_{ij}\right]{\eta}_{j}^{+}\nonumber \\
&-&\left[D\left(1+{\kappa^2}/{q^2}\right) q_i q_j G^-_{\rho}+i D {\cal E}_i q_j G^-_{c}-\delta_{ij}\right]{\eta}_{j}^{-}, \nonumber \\ \label{eqn:jrho} 
\end{eqnarray}
and
\begin{eqnarray}
J_{c i}&=&\left[D q_i q_j G^+_{c}-i D {\cal E}_i q_j G^+_{\rho}-\delta_{ij}\right]{\eta}_{j}^{+}\nonumber \\
&+&\left[D q_i q_j G^-_{c}+i D {\cal E}_i q_j G^-_{\rho}-\delta_{ij}\right]{\eta}_{j}^{-}. \label{eqn:jc} 
\end{eqnarray}
Note that we have ignored the DC component of the current density in equation (\ref{eqn:currentd}).

We can now complete the calculation of the fluctuations by specifying the spectrum of the noise terms. Since the noise originates from Poissonian number fluctuations, a consistent prescription within our calculations will be to regard them as independent Gaussian distributed white noise terms of zero mean, with variances that are controlled by the average concentration, namely
\begin{equation} \label{eq:noise-real-space}
\langle \eta^a_{i}\left(\myvec{x},t\right) \eta^{b}_{j}(\myvec{x}',t')\rangle = 2D C_{\textrm{pore}} \delta^{ab} \delta_{ij} \delta^3(\myvec{x}-\myvec{x}') \delta(t-t'), \\
\end{equation}
in real space, and
 \begin{eqnarray} \label{eq:noise-fourier}
\langle {\eta}^a_{i}(\myvec{q},\omega){\eta}^{b}_{j}(\myvec{q}',\omega')\rangle &=& 2D C_{\textrm{pore}} \delta^{ab} \delta_{ij} \nonumber \\
& \times & (2\pi)^{4} \delta^3(\myvec{q}+\myvec{q}') \delta(\omega+\omega'),
\end{eqnarray}
in Fourier space, where $a$ and $b$ can be $+$ or $-$.

Performing the averaging over the noise terms, we obtain the following expression for the number density (concentration) fluctuations
\begin{equation}
\langle|c({\bm{q}},\omega)|^2\rangle= 2D C_{\textrm{pore}} q^2 \left(|G^+_{c}|^2+|G^-_{c}|^2\right),\label{eq:cc-res}
\end{equation}
and the following expression for the charge density fluctuations
\begin{equation}
\langle|\rho({\bm{q}},\omega)|^2\rangle= 2D C_{\textrm{pore}} q^2 \left(|G^+_{\rho}|^2+|G^-_{\rho}|^2\right).\label{eq:rhorho-res}
\end{equation}
For the fluctuations of the fluxes, we obtain
\begin{strip}
\begin{eqnarray}
&&\langle|\bm{J}_\rho({\bm{q}},\omega)|^2\rangle= 2D C_{\textrm{pore}}\nonumber \\
&&\hskip0.4cm \times  \Big[6+ D^2 (q^2+\kappa^2)^2 \left(|G^+_{\rho}|^2+|G^-_{\rho}|^2\right)+D^2 q^2 {\cal E}^2  \left(|G^+_{c}|^2+|G^-_{c}|^2\right)-D^2 (q^2+\kappa^2) \left(G^+_{\rho}+G^{+*}_{\rho}+G^-_{\rho}+G^{-*}_{\rho}\right) \nonumber \\
&& \hskip0.7cm + i D (\bm{q} \cdot {\cal E}) \left(G^+_{c}-G^{+*}_{c}-G^-_{c}+G^{-*}_{c}\right) + i D^2 (q^2+\kappa^2)  (\bm{q} \cdot {\cal E}) \left(G^+_{\rho} G^{+*}_{c}-G^{+*}_{\rho} G^+_{c}+G^{-*}_{\rho}G^-_{c}-G^-_{\rho} G^{-*}_{c}\right) \Big],\label{eq:Jrho2}
\end{eqnarray}
and
\begin{eqnarray}
&&\langle|\bm{J}_c({\bm{q}},\omega)|^2\rangle= 2D C_{\textrm{pore}}\nonumber \\
&&\hskip0.4cm \times  \Big[6+ D^2 q^4 \left(|G^+_{c}|^2+|G^-_{c}|^2\right)+D^2 q^2 {\cal E}^2  \left(|G^+_{\rho}|^2+|G^-_{\rho}|^2\right)-D^2 q^2 \left(G^+_{c}+G^{+*}_{c}+G^-_{c}+G^{-*}_{c}\right) \nonumber \\
&& \hskip0.7cm + i D  (\bm{q} \cdot {\cal E}) \left(G^+_{\rho}-G^{+*}_{\rho}-G^-_{\rho}+G^{-*}_{\rho}\right) - i D^2 q^2  (\bm{q} \cdot {\cal E}) \left(G^+_{\rho} G^{+*}_{c}-G^{+*}_{\rho} G^+_{c}+G^{-*}_{\rho}G^-_{c}-G^-_{\rho} G^{-*}_{c}\right) \Big].\label{eq:Jc2}
\end{eqnarray}
\end{strip}
\subsection{Approximating the Geometry of the System}\label{sec:fourier-apprx-scheme}

\begin{figure}[b]
\centering
\includegraphics[width=1.0\columnwidth]{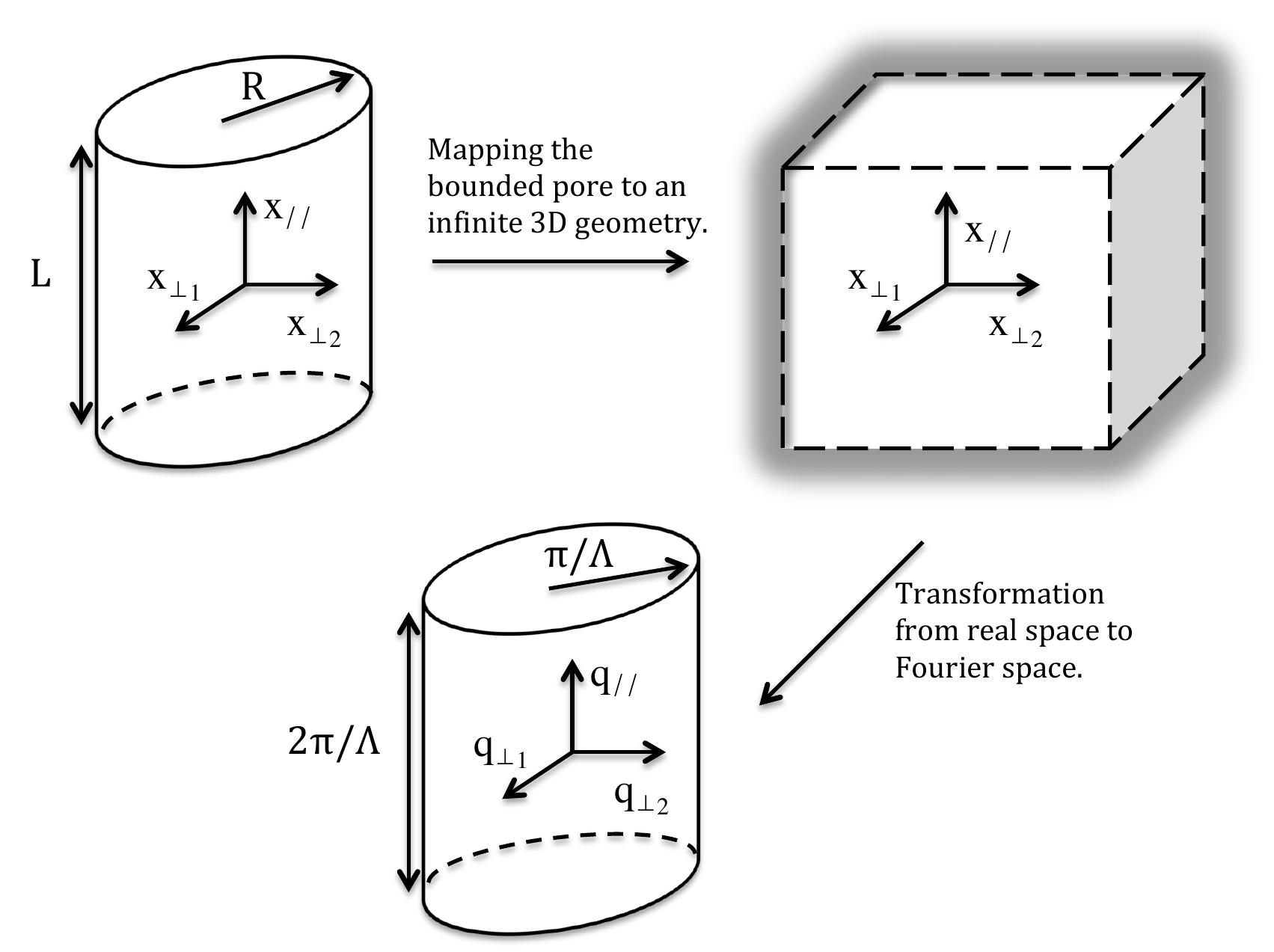}
\caption{A schematic drawing illustrating the transformation from a bounded cylindrical
geometry to an infinite 3D space then to a confined cylindrical domain in Fourier space.}
\label{fig:fourier-scheme}
\end{figure}

Since we are interested in current fluctuations in a pore of a given geometry, a complete description of the problem will require solution of the above stochastic field equations subject to the specific boundary conditions imposed by the geometry of the pore. Instead of attempting this route, we use an {\em ad hoc} approximation scheme in which we first assume that the system admits translation invariance in all directions, which means that it can be solved using Fourier transformation. Then, when integrating over the Fourier modes, we impose infrared and ultraviolet cutoffs on the wave-vectors using the corresponding geometric characteristics along each direction. This scheme should capture the essence of the finite size of the pore, and give the right trends when we investigate the effect of the pore dimensions. This approximation scheme is pictorially described in figure \ref{fig:fourier-scheme}. In the resulting geometry in Fourier space, we describe the coordinates of our system using the wave vector ${\bm{q}}=(q_{\perp1}, q_{\perp2},q_{\parallel})$.  We choose $q_{\parallel}$ to correspond to the parallel direction along the pore length $L$, whereas $q_{\perp1},$ and $q_{\perp2}$ are used to describe the perpendicular directions, characterized in position space by a radius $R$ as shown in figure {\ref{fig:fourier-scheme}}. We also choose the applied electric field inside the pore to be in the parallel direction, i.e. ${\bm{\mathcal{E}}}= (0,0,{\mathcal{E}})$. 

The quantity of interest is the electric current that flows across the ion channel. Therefore, we need to calculate the parallel component of the current density fluctuations. We find
\begin{strip}
\begin{eqnarray}
\hskip 1cm &&\left\langle \left|{J}_{\rho\parallel}\left({\bm{q}},\omega\right)\right|^2 \right\rangle=4DC_{{\rm{pore}}}\Bigg\{1+\dfrac{D^4 \mathcal{E}^2 [\kappa^4 q_{\parallel}^4+q^4 (q_{\perp}^4-2 q_{\parallel}^4)]-D^4 q^4 q_{\parallel}^2 (\kappa^2+q^2)^2}{q^2\big[(-\omega^2+D^2q^4+D^2q^2\kappa^2+D^2q_{\parallel}^2\mathcal{E}^2)^2+\omega^2 D^2 (2 q^2+\kappa^2)^2\big]} \nonumber \\
&&\hskip1cm +\dfrac{\omega^2 \big[ D^2\mathcal{E}^2q^2(2q_{\parallel}^2+q^2)-D^2 q_{\parallel}^2(\kappa^2+q^2)^2 \big]+ D^4 \mathcal{E}^2 (q_{\perp}^2-q_{\parallel}^2) [\mathcal{E}^2 q_{\parallel}^2q^2+ \kappa^2 q^2 (2 q^2+\kappa^2)]}{q^2\big[(-\omega^2+D^2q^4+D^2q^2\kappa^2+D^2q_{\parallel}^2\mathcal{E}^2)^2+\omega^2 D^2 (2 q^2+\kappa^2)^2\big]}\Bigg\}.\label{eqn:current12}
\end{eqnarray}
\end{strip}
\noindent We proceed to determine the fluctuations in the current that passes through the pore, which we define as the normal charge density flux that crosses the pore of area $A$ averaged over the length of the pore, namely
\begin{equation}
I_{\rm ave}(t)\equiv \frac{1}{L}\int \rmd x_{\parallel} \int_A \rmd x_{\perp1} \rmd x_{\perp2}\, J_{\rho \parallel}(\bm{x},t).
\end{equation}
The current fluctuations, $S(\omega)$, is defined as follows
\begin{equation}
S(\omega)=\langle \left| I_{\rm ave}(\omega) \right|^2\rangle,
\end{equation}
and can be evaluated in terms of the above expression for the charge density flux fluctuations and some geometric factors that enter through the limits of integration over the wave-vectors (see figure \ref{fig:fourier-scheme}) as follows
\begin{eqnarray} 
S(\omega) &=& \frac{1}{2 \pi^2}  \int^{\pi\Lambda^{-1}}_{\pi L^{-1}} \rmd q_{\parallel}\int^{\pi\Lambda^{-1}}_{\pi R^{-1}} q_{\perp} \rmd q_{\perp} \nonumber \\
&& \times \big|{\cal A}(q_{\perp})\big|^2\big|{\cal L}(q_{\parallel})\big|^2 \left\langle \left|{J}_{\rho\parallel}\left({\bm{q}},\omega\right)\right|^2 \right\rangle.\label{eqn:psd-derivation}
\end{eqnarray}
where ${\cal{A}}= \int_A \rmd x_{\perp1} \rmd x_{\perp2}\;e^{i q_{\perp1} x_{\perp1}+i q_{\perp2} x_{\perp2}}$ and ${\cal L}= \frac{1}{L}\int \rmd x_{\parallel}\;e^{i q_{\parallel} x_{\parallel}}$. Here, the lateral area form factor
\begin{equation}\label{eqn:area-cylinder}
\big|{\cal{A}}(q_{\perp})\big|^2=\left(\pi R^2\right)^2 \left[\frac{2 J_{1}\left(q_{\perp} R\right)}{q_{\perp} R}\right]^2,
\end{equation}
and the longitudinal form factor
\begin{equation}
\big|{\cal{L}}(q_{\parallel})\big|^2 =\dfrac{2}{(L q_{\parallel})^{2}}  \big[1 - \cos(Lq_{\parallel})\big].\label{eqn:length}
\end{equation}
encode information about the geometric features of the pore. While the exact form of these functions will depend on the specific geometric features, the overall magnitude of the functions will be controlled by the characteristic length scales. Therefore, we expect the approximation scheme that is sketched in figure \ref{fig:fourier-scheme} to only cause a numerical error of order unity. It is also worth mentioning that equation (\ref{eqn:psd-derivation}) can be applied to the case where we measure the current over a slice through the pore, instead of performing a length-averaging, in which case we need to set ${\cal{L}}(q_{\parallel})=1$. This will not lead to qualitative changes in the behaviour of the spectrum.

\begin{figure}[t]
\centering
\includegraphics[width=1.0\columnwidth]{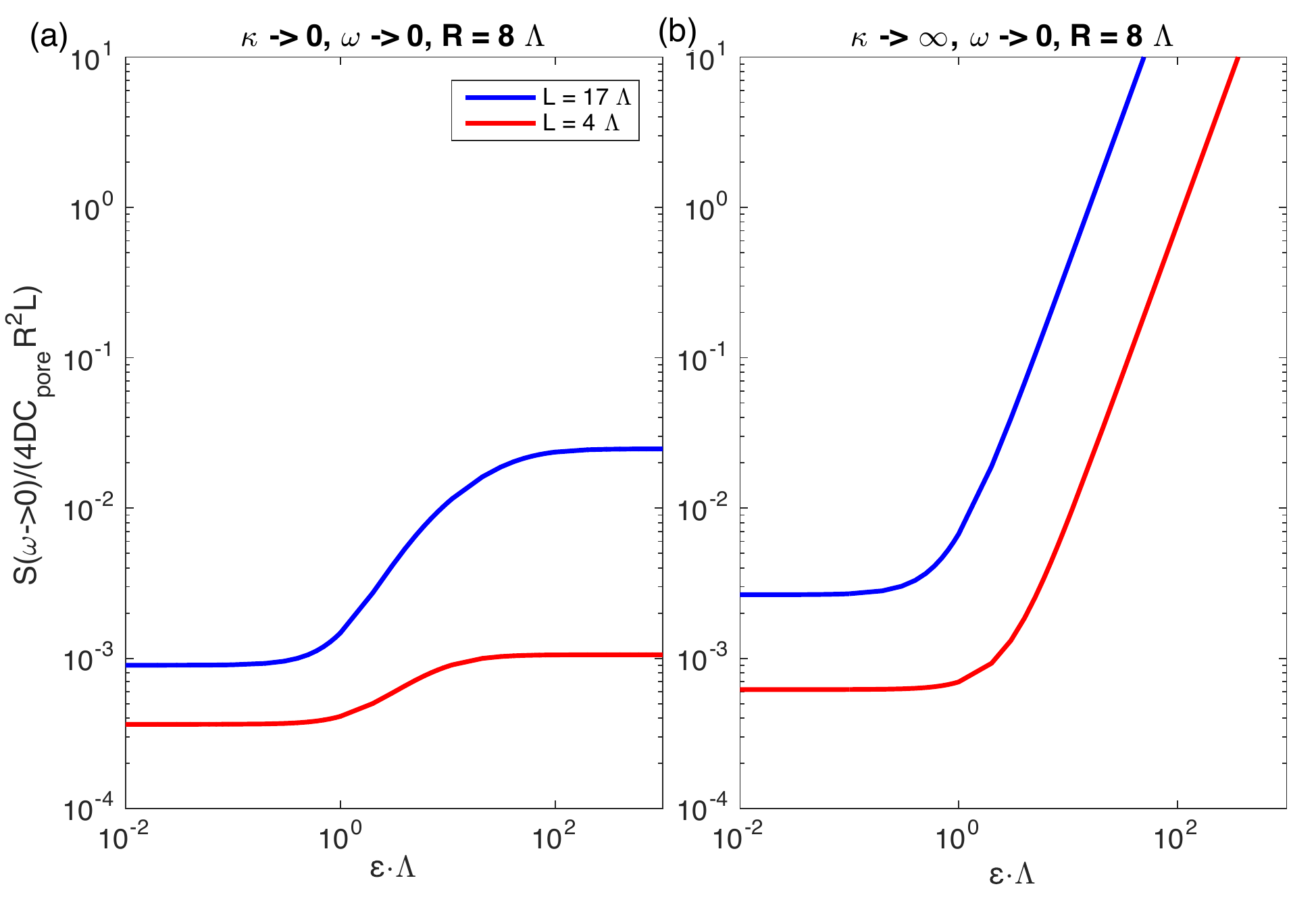}
\caption[Power spectral density at low frequencies]{
(a) The current fluctuations as $\omega \to 0$ as a function of the applied electric field, with $\kappa$ approaching $0$. (b) The current fluctuations as $\omega \to 0$ as a function of the applied electric field, with $\kappa$ approaching infinity.}
\label{fig:omegazero}
\end{figure}

\subsection{Examining the Different Asymptotic Forms}\label{sec:asymp_beh}

From the expression of the flux density fluctuations given in equation (\ref{eqn:current12}), we observe that current fluctuations will depend---through competing contributions---on the electric field, the geometric features of the pore, and the Debye length. To develop a physical intuition about the behaviour of $S(\omega)$ at different frequencies as a function of all these parameters, we examine the different asymptotic limits of equation (\ref{eqn:current12}).

The generic form of equation (\ref{eqn:current12}) as a function of $\omega$ is a ratio between two fourth order polynomials. At high frequencies, the expression has a plateau that is determined by the Gaussian white noise in the system
\begin{equation}
 \lim_{\omega \to \infty}\left\langle \left|{J}_{\rho\parallel}\left({\bm{q}},\omega\right)\right|^2 \right\rangle= 4 D C_{{\rm{pore}}}.
\end{equation}
Since the polynomial expressions are of the same order, the asymptotic limit of equation (\ref{eqn:current12}) at low frequencies is also finite, but with much richer structure than the high frequency limit:
\begin{eqnarray}
&&\left\langle \left|{J}_{\rho\parallel}\left({\bm{q}},0\right)\right|^2 \right\rangle=\frac{ 4 D C_{{\rm{pore}}}  \; q_{\perp}^2}{q^2\big[(q^4+q^2\kappa^2+2q_{\parallel}^2\mathcal{E}^2)^2\big]}\left\{\mathcal{E}^4 q_{\parallel}^2 q^2\right.
 \nonumber \\
&&\hskip 0.3cm\left.+\mathcal{E}^2 (q^2+\kappa^2)\big[\kappa^2 q_{\perp}^2+q^2 (q_{\parallel}^2+q^2)\big] +q^4 (q^2+\kappa^2)^2 \right\},\nonumber \\
\label{eq:limit0}
\end{eqnarray}
As $\kappa \to 0$, equation (\ref{eq:limit0}) reduces to an equation that depends on the applied electric field and the pore dimensions:
\begin{equation}\label{eq:lowkappa}
\lim_{\kappa\to 0}\left\langle \left|{J}_{\rho\parallel}\left({\bm{q}},0\right)\right|^2 \right\rangle=4DC_{{\rm{pore}}}\dfrac{q_{\perp}^2(q_{\parallel}^2+q_{\perp}^2+\mathcal{E}^2)}{\big[(q_{\parallel}^2+q_{\perp}^2)^2+q_{\parallel}^2\mathcal{E}^2 \big]}. 
\end{equation}
On the one hand, at high values of the electric field, the magnitude of $S(\omega)$ at low frequencies will be controlled by $(q_{\perp}/q_{\parallel})^2$, which depends on the aspect ratio of the pore. On the other hand, at lower values of the electric field, a systematically smaller amplitude of the noise is obtained (as compared with the high frequency limit), which becomes less significant when the length $L$ of the channel approaches its radius $R$ as shown in figure \ref{fig:omegazero}. At high values of $\kappa$, a continuously increasing behaviour is observed at high values of $\mathcal{E}$ as indicated in figure \ref{fig:omegazero}(b). However, at low values of the electric field, $S(\omega)$ has a plateau. 

At high and low electric field limits, all terms involving $\kappa$ in equation (\ref{eq:limit0}) cancel and we obtain the same behaviour as we found previously from equation (\ref{eq:lowkappa}), namely
\begin{subequations}\label{eq:low_e_high_e}
\begin{align}
\lim_{\mathcal{E}\to 0}\left\langle \left|{J}_{\rho\parallel}\left({\bm{q}},0\right)\right|^2 \right\rangle&= 4DC_{{\rm{pore}}}\;\dfrac{q_{\perp}^2}{(q_{\parallel}^2+q_{\perp}^2)},\\
\lim_{\mathcal{E}\to\infty}\left\langle \left|{J}_{\rho\parallel}\left({\bm{q}},0\right)\right|^2 \right\rangle &= 4DC_{{\rm{pore}}}\;\dfrac{q_{\perp}^2}{q_{\parallel}^2}.
\end{align}
\end{subequations}
\begin{figure}[t]
\centering
\includegraphics[width=1.0\columnwidth]{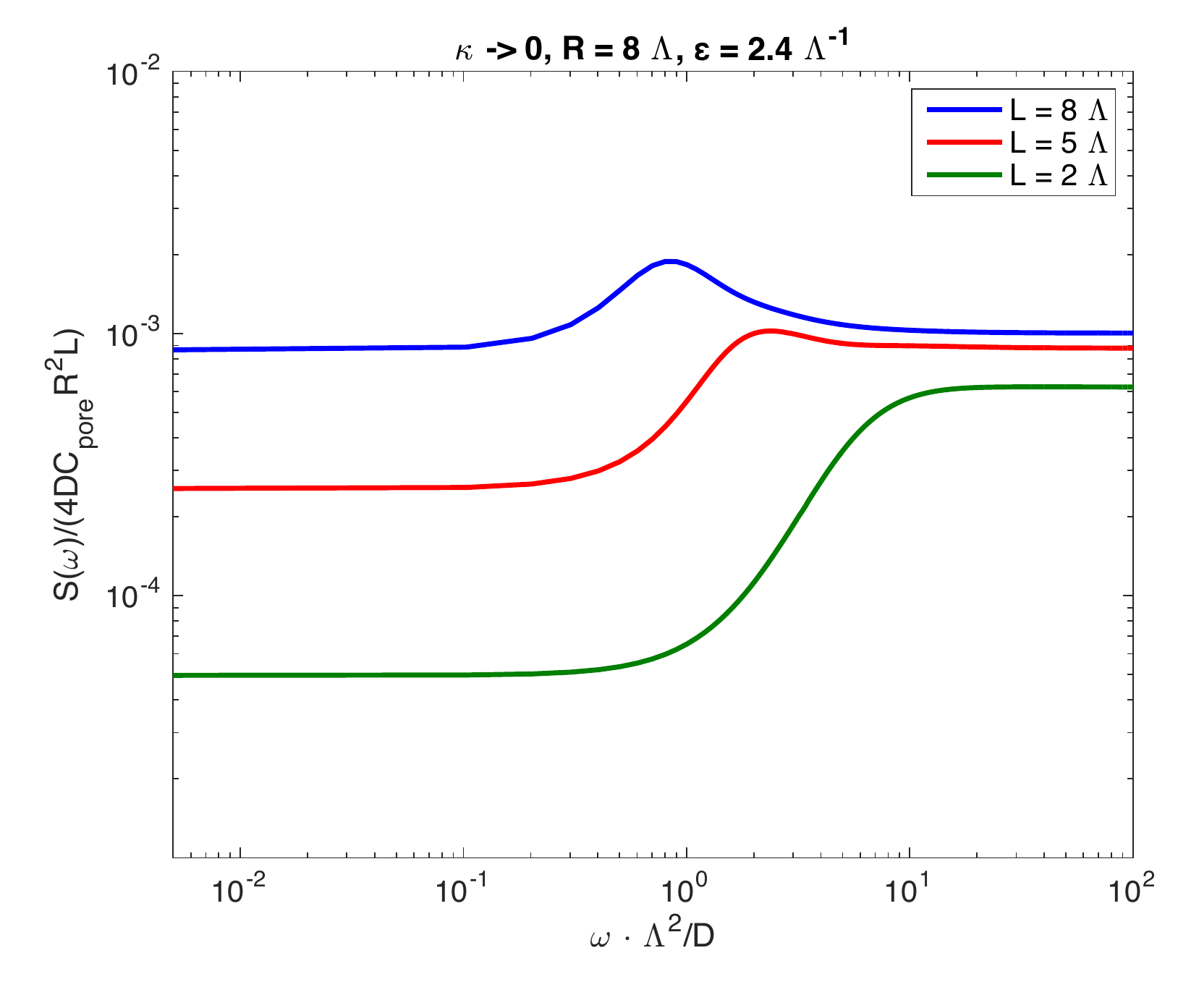}
\caption[Current fluctuations at high values of the Debye length]{
The current fluctuations as $\kappa \to 0$ as a function of the frequency $\omega$, using pores with radius $R=8 \varLambda$ and lengths $L= 1.5,\; 4,\; 8 \varLambda$.
}
\label{fig:lowkappa}
\end{figure}

Let us now consider the behaviour of $S(\omega)$ at low and high values of the inverse Debye length $\kappa$ as a function of the frequency $\omega$:
\begin{eqnarray}
&& \lim_{\kappa\to 0}\left\langle \left|{J}_{\rho\parallel}\left({\bm{q}},\omega\right)\right|^2 \right\rangle= 4 D C_{{\rm{pore}}} \nonumber \\ 
&& \hskip0.1cm\times\dfrac{\big[\omega^2+D^2\mathcal{E}^2q_{\parallel}^2+D^2 q^4\big] \big[\omega^2+D^2q_{\perp}^2(q^2+\mathcal{E}^2)\big]}{\omega^4+2D^2\omega^2\big[q^4-\mathcal{E}^2q_{\parallel}^2\big]+D^4\big[q^4+q_{\parallel}^2\mathcal{E}^2\big]^2}.\label{eq:lowkappa2}
\end{eqnarray}
Under this condition, $S(\omega)$ exhibits a plateau at low frequencies and a decreasing power law at high frequencies followed by the white noise. However, if we consider a pore with a radius considerably larger than its length a different behaviour is observed; under this condition, the shape of the function reverts, showing a weaker plateau at small frequencies followed by an increasing power law as shown in figure \ref{fig:lowkappa} above. 

\begin{figure}[t]
\centering
\includegraphics[width=1.0\columnwidth]{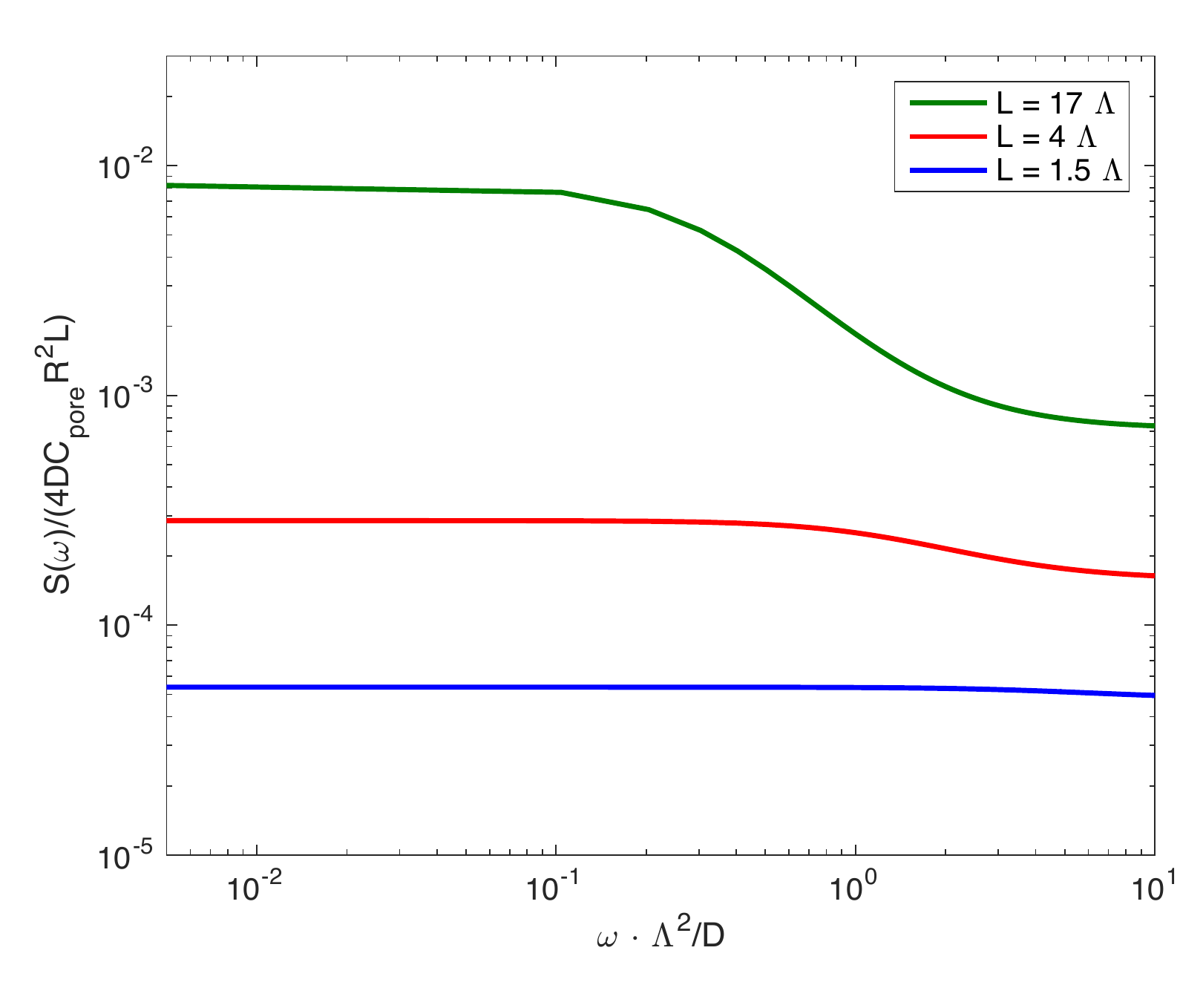}
\caption[Current fluctuations at low Debye length limit]{
The current fluctuations as $\kappa \to \infty$ as a function of the frequency $\omega$ in a pore with radius $R=8 \varLambda$~and lengths $L=1.5, \; 4, \;17 \varLambda$. 
}
\label{fig:shortpores}
\end{figure}

In the limit where $\kappa$ approaches infinity, equation (\ref{eqn:current12}) yields:
\begin{eqnarray}
&& \lim_{\kappa\to\infty}\left\langle \left|{J}_{\rho\parallel}\left({\bm{q}},\omega\right)\right|^2 \right\rangle=4 D C_{{\rm{pore}}} \nonumber \\
&& \hskip1.5cm \times \dfrac{q_{\perp}^2 \big[\omega^2+D^2\mathcal{E}^2q_{\perp}^2+D^2(q_{\parallel}^2+q_{\perp}^2)^2 \big]}{(q_{\parallel}^2+q_{\perp}^2) \big[\omega^2+D^2(q_{\parallel}^2+q_{\perp}^2)^2\big]}.\label{eq:highkappa2}
\end{eqnarray}
In this limit, $S(\omega)$ does not exhibit any peculiar behaviour, except when dealing with a very short pore as shown in figure \ref{fig:shortpores}. In this case, $S(\omega)$ is almost flat as the difference between the low frequency plateau and the high frequency white noise becomes negligible.

\begin{figure}[b]
\centering
\includegraphics[width=1.0\columnwidth]{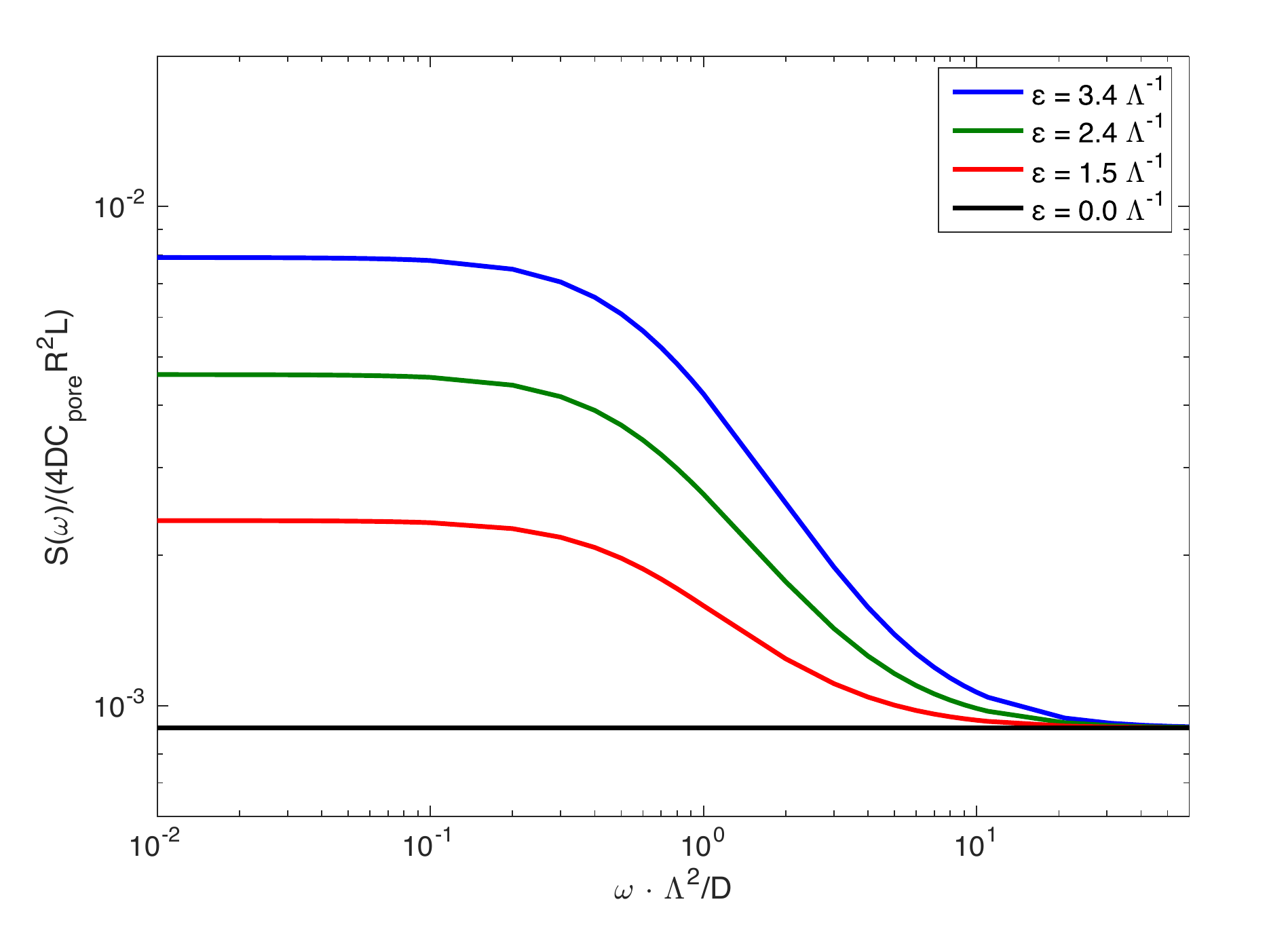}
\caption[Current fluctuations at high Debye length limit and different electric fields]{
The current fluctuations as $\kappa \to \infty$ as a function of the frequency $\omega$ in a pore with radius $R=8 \varLambda$~ and length $L=17 \varLambda$. Four different values of the electric field are used $\mathcal{E}=0.0,\; 1.5,\; 2.4,\; 3.4 {{\varLambda}}^{-1}$. 
}
\label{fig:electric}
\end{figure}

\begin{figure}[t]
\centering
\includegraphics[width=1.0\columnwidth]{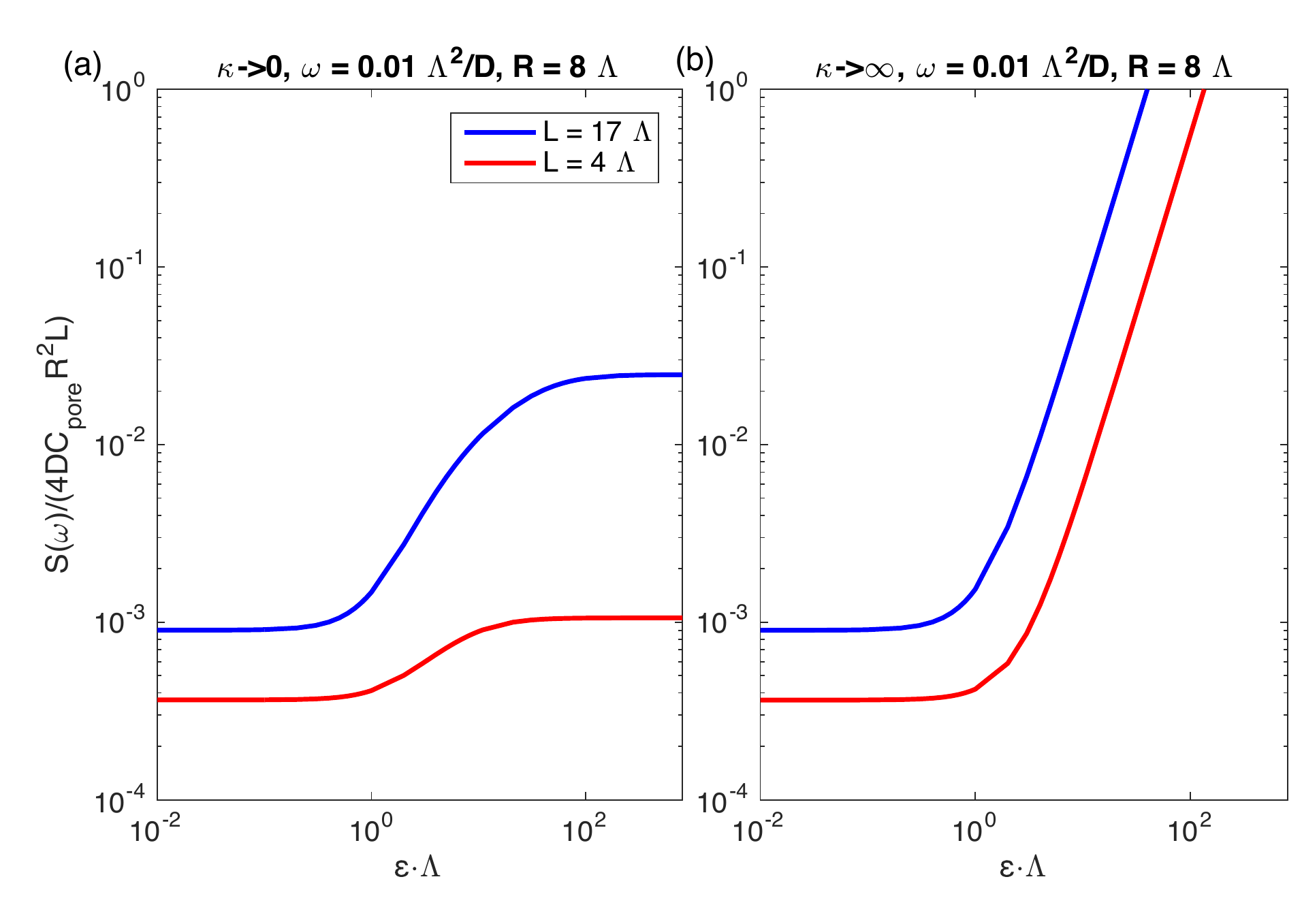}
\caption[Current fluctuations at a fixed frequency as a function of electric field]{
(a) The current fluctuations as $\kappa \to 0$ as a function of the applied electric field, at a fixed frequency $\omega = 0.01 {{\varLambda}}^2/D$. (b) The current fluctuations as $\kappa \to \infty$ as a function of the applied electric field, at a fixed frequency $\omega = 0.01 {{\varLambda}}^2/D$.
}
\label{fig:debye_length}
\end{figure}

Figure \ref{fig:electric} illustrates the behaviour of equation (\ref{eq:highkappa2}) at different values of the electric field. As figure \ref{fig:electric} indicates, the spectrum of current fluctuations becomes frequency independent as we decrease the magnitude of the applied electric field. This highlights the major role played by the electric field in generating the power law at high frequencies. To elaborate more on this behaviour, we plot the current fluctuations at a certain specified low frequency, e.g. $\omega = 0.01 {{\varLambda}}^2/D$, as a function of the applied electric field; see figure \ref{fig:debye_length}. We find that the current fluctuations will always exhibit a higher amplitude at high values of the electric field, a result that becomes more significant when dealing with longer pores.

We now study the low electric field limit. Under this condition, the expression of the current density fluctuations reduces to:
\begin{equation}
\lim_{\mathcal{E}\to 0}\left\langle \left|{J}_{\rho\parallel}\left({\bm{q}},\omega\right)\right|^2 \right\rangle=4DC_{{\rm{pore}}}\dfrac{\omega^2 q^2+D^2 q_{\perp}^2 (\kappa^2+q^2)^2}{q^2 \big[\omega^2+D^2(\kappa^2+q^2)^2\big]}.
\end{equation}
The limit at low frequencies was developed previously in equation (\ref{eq:low_e_high_e}a), highlighting the existence of a structure only when dealing with very long pores. However, at this limit, the role of the Debye length becomes negligible highlighting once more the major role played by the electric field in generating the frequency dependent behaviour.

\begin{figure*}
\centering
\includegraphics[width=1.8\columnwidth]{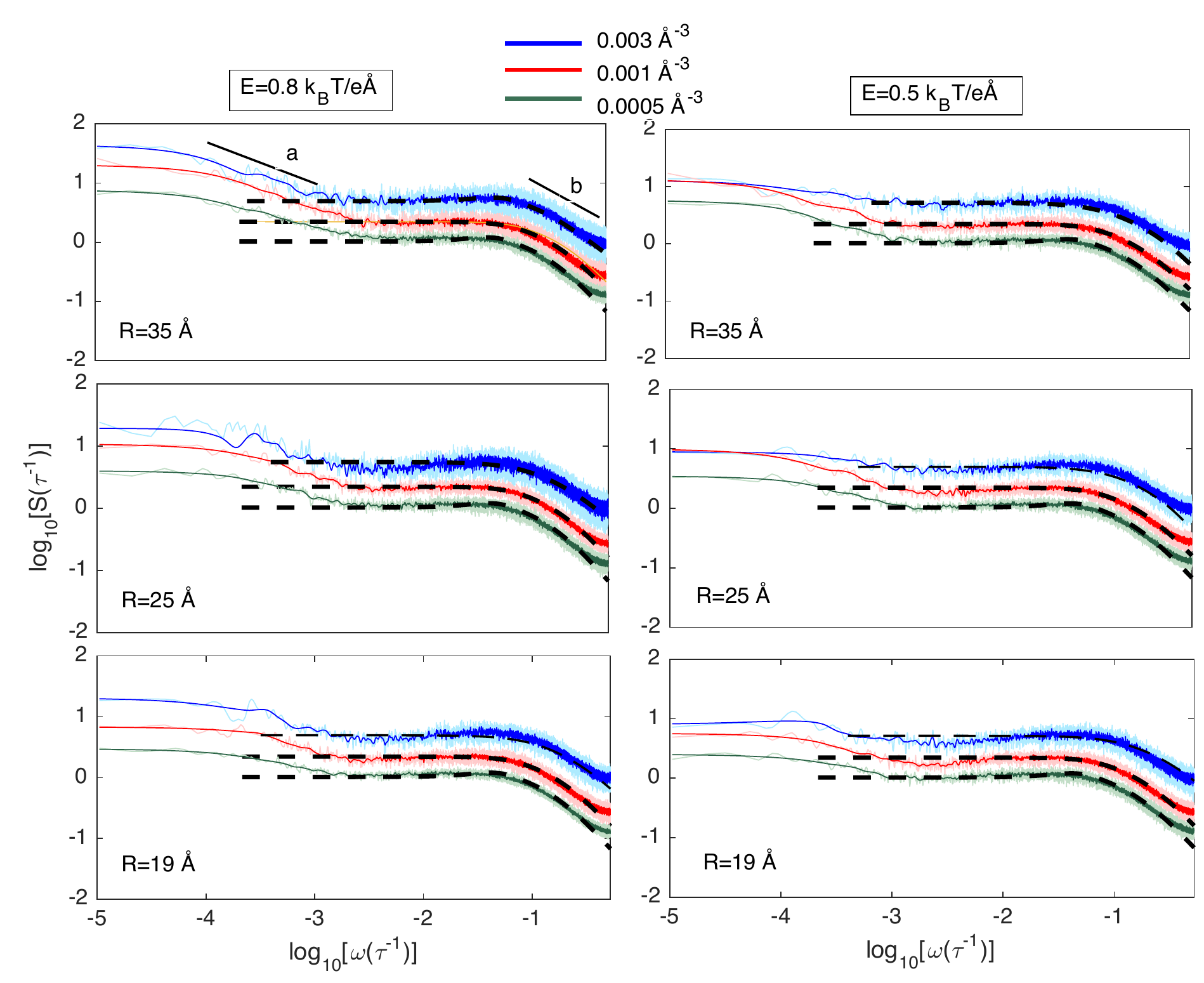}
\caption[Simulation results: current fluctuations extracted from simulations with various ion concentrations]{
The current fluctuations $S\left(\omega\right)$ in units of the inverse of the time scale $\tau$, as a function of the frequency $\omega$ on a log-log scale, for ion concentrations (a) $C_0 = 5\times 10^{-4}$~\AA$^{-3}$ (0.8 mol/l), (b) $C_{0}  = 1\times 10^{-3}$~\AA$^{-3}$ (1.7 mol/l), and (c) $C_{0}  = 3\times 10^{-3}$~\AA$^{-3}$ (5.0 mol/l). The solid curves denote $S\left(\omega\right)$ from simulations, smoothed using a moving average for improved clarity, and the shaded areas indicate the standard deviations of the blocks. The dashed lines denote fits with equation (\ref{eqn:psd-derivation}). The diffusion coefficient $D = 1$~\AA$^2/\tau$ and the small-scale cutoff length is set to the ion size, $\Lambda = 3$~\AA, for all curves.
The applied electric field are $E=0.5 \; \kT$/($e$\AA) and $E=0.8 \;\kT$/($e$\AA).
}
\label{fig:concentration}
\end{figure*}
\section{Brownian Dynamics Simulations}\label{sec:simulation}
To check the validity of our theoretical description, we perform coarse-grained Langevin dynamics simulations using the many-particle package Espresso {\cite{2006_Limbach}} to derive the current fluctuations in an ion channel. The Langevin equation of the particle $i$ reads:
\begin{equation} \label{eqn:langevin}
m_{i} \ddot{\bm{r}}_{i} = -\sum_{j\neq i} \nabla V_{ij} + \myvec{F}_{i} - {\gamma} \dot{\bm{r}}_{i} + \myvec{\xi}_{i},
\end{equation}
where $m_i$ and $\bm{r}_i$ denote the particle mass and position, respectively, while $V_{ij}$ and $\bm{F}_i$ represent particle interactions and external forces. The random force $\myvec{\xi}_{i}$ is characterized by zero mean, and a variance $\langle \myvec{\xi}_{\textit{i}}\left(t\right) \cdot \myvec{\xi}_{\textit{j}}\left(t'\right)\rangle = 6 \kT \gamma \delta_{\textit{ij}} \delta\left(t-t'\right)$, whereas $\gamma = 1$~$\kT \tau$/\AA$^2$ is set to be the friction coefficient with $\tau$ being the simulation time scale. For simplicity, we consider all the particles to possess the same mass. The diffusion coefficient is constant with a value of $D=1 \;{\rm{\AA}}^2/\tau$.

We build an artificial membrane consisting of frozen particles on a cubic lattice with a lattice constant of $6$~\AA, of width $W=100$~\AA, length $L=48$~\AA, and height  $H=48$~\AA (see figure \ref{fig:setup}). A cylindrical pore is created through the membrane by removing particles covering a specified radius that we vary between $19$, $25$, and $30$~\AA. 
Two reservoirs of randomly distributed ions with equal concentrations $C_0$ are added to both sides of the pore with a height $H=50~{\rm{\AA}}$. The ions will interact with each other electrostatically, but they will also experience a Weeks-Chandler-Anderson potential:
\begin{equation} \label{eqn:potential}
V_{ij}= \ell_{\rm B}\kT\frac{Q_{i}Q_{j}}{r_{ij}} + 4 \epsilon_{ij} \left[ \left(\frac{\sigma_{ij}}{r_{ij}}\right)^{12}-\left(\frac{\sigma_{ij}}{r_{ij}}\right)^{6}\right] + V_{\textrm{shift}},
\end{equation}
where $Q_i=\pm 1$ being the $i$-th particle charge, and $r_{ij}$ denoting the distance between the $i$-th and $j$-th particles. The repulsive potential is truncated at $r_{ij} = 2^{1/6}\sigma_{ij}$, where instead we use a constant value $V_{ij}=\epsilon_{ij}/4$. The concentration of the ions $C_0$ takes three different values: $C_{0} = \{3\times 10^{-3},1\times 10^{-3},5\times 10^{-4}\}$~\AA$^{-3}$, corresponding to $C_{0} = \{0.8,~1.7,~5.0\}$~mol/l. Each ion crossing the pore, will experience a constant external force along the $x_{\parallel}$ direction:
\begin{equation} \label{eq:external-force}
\myvec{F}_{i} = \Bigg\{ {\begin{array}{l l} 
Q_{i} \myvec{E} & \quad \text{if $0<x_{\parallel}<L$ }\\
 0 & \quad \text{otherwise},
  \end{array}}
\end{equation}
The electric field is defined as $\myvec{E}=(0,0,E)$, and $E$ is set to $0.5$ and $0.8$~$\kT/\left(e\rm{\AA}\right)$ equivalent to a potential difference of $0.6$~V and $1.0$~V across the pore. Using a time step of $0.006\;\tau$, we perform simulations of $4\times 10^8$~$\tau$, where the simulation time scale $\tau$ is calculated to be $5$ ps. Every $100$ steps, the velocities of the ions that cross the pore are used to calculate the current through the membrane as follows:
\begin{equation}
I(t)=\frac{1}{L}\sum_{i}q_{i} {\bm{u_i}}.
\end{equation}
By applying Welch's method over the collected data, we calculate the current fluctuations in frequency domain $S_{I}(\omega)=\big|I(\omega) \big|^2$.

{\begin{figure}[t]
\centering
\includegraphics[width=1.0\columnwidth]{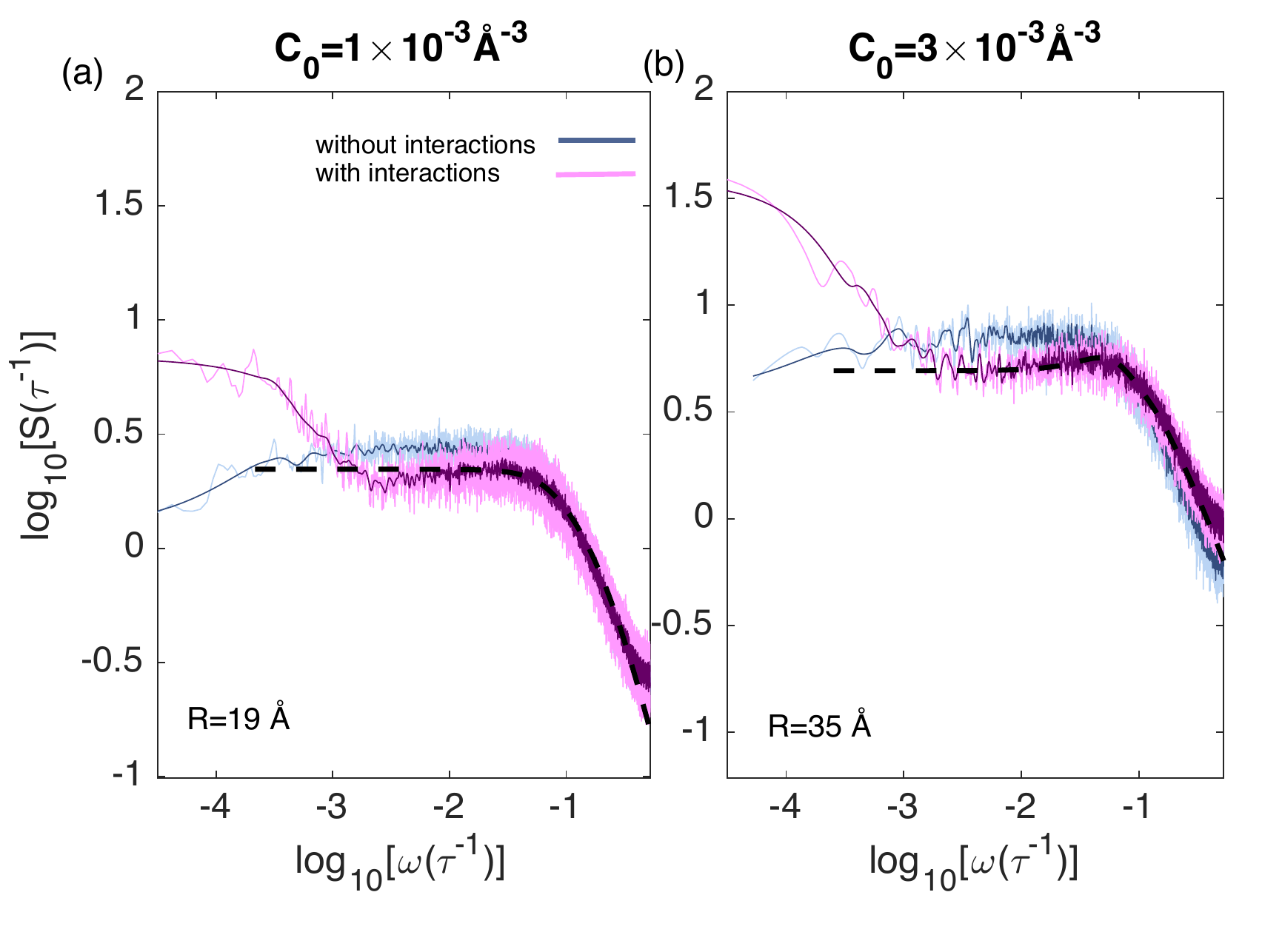}
\caption[Influence of ion-ion interactions on the current fluctuations]{
The spectrum of current fluctuations with the ion-ion interaction potential $V_{ion-ion}\left(r_{ij}\right)$ given by equation \ref{eqn:potential} (with interactions, violet line) and with $V_{ion-ion}\left(r_{ij}\right) = 0$ (without interactions, blue line).
The spectrum on the left panel is calculated in a channel with $R = 19$~\AA, $E = 0.8$~$\kT/(e\textrm{\AA})$, and $C_{0}=1\times 10^{-3}$~\AA$^{-3}$. The spectrum on the right panel is calculated in a channel with $R = 35$~\AA, $E = 0.8$~$\kT/(e\textrm{\AA})$, and $C_{0}=3\times 10^{-3}$~\AA$^{-3}$.
}
\label{fig:interactions}
\end{figure}}

In figure \ref{fig:concentration}, we show the obtained spectrum of ionic current fluctuations from the simulations performed over a range of various parameters. The power law behaviour observed at low frequencies has a slope that increases with concentrations, radius, and electric field, reaching $1/\omega^{0.62}$ at the highest concentration. This behaviour is consistent with our previous reports {\cite{zorkot2016-1,zorkot2016-2}}. The black dashed lines represent the theoretical prediction of equation (\ref{eqn:psd-derivation}), which provide perfect fits for all values of the parameters used without any further rescaling.

To elaborate more on the role of ion-ion correlation that was highlighted as a major source of the power law at low frequencies {\cite{zorkot2016-1,zorkot2016-2}}, and to further check the validity of our theoretical description, we run simulations with $V_{ion-ion}$ set to $0$; the corresponding spectra are shown in figure {\ref{fig:interactions}}. We use two pores with radii $R = 19$~\AA~ and $R = 25$~\AA~. The ions concentration are set to $C_{0}=1\times 10^{-3}$~\AA$^{-3}$ and $C_{0}=3\times 10^{-3}$~\AA$^{-3}$ with a constant applied electric field  $E= 0.8$~$\kT/(e\textrm{\AA})$. Without interactions, the power law at low frequencies is replaced by a plateau; a behaviour that matches quantitatively with our theoretical predictions (dashed lines) without any further adjustments regardless of the parameters used. 

\section{Simplified Linearized Theory}\label{sec:simple_theory}

As we have mentioned above, the linearized theory of current fluctuations presented in section {\ref{sec:theory}} matches with the results of our simulations without any fitting or adjustment. Therefore, it provides a comprehensive description of the system despite the {\em ad hoc} approximations made about the geometry and use of Fourier modes depicted in figure \ref{fig:fourier-scheme}. The resulting expression for the parallel electric current flux density fluctuations given in equation (\ref{eqn:current12}) is, however, quite involved, and it will be helpful if simpler approximate expressions can be extracted without losing the essence of the theory.

The first simplification could arise if the size of the pore is sufficiently small as compared to the Debye length. In this case, we can apply the $\kappa \to 0$ limit and use the simpler expression given in equation (\ref{eq:lowkappa2}). In our previous publications {\cite{zorkot2016-1,zorkot2016-2}}, we have presented an even simpler expression in the form of
\begin{eqnarray}
&&\langle|{J}_{\rho\parallel}({\bm{q}},\omega)|^2\rangle \approx 4DC_{{\rm{pore}}}\nonumber \\
&& \hskip0.5cm \times \left[1+\frac{{\mathcal{E}}^2  q^2 ({\omega^2}{D^2} + D^4 q_{\perp}^{4})} 
       {(D^2 q^4+ D^2{ \mathcal{E} }^2 q_{\parallel}^2 - {\omega^2})^2+ 4 {\omega^2}{D^2} q^4}\right], \label{eqn:simplified}
\end{eqnarray}
which we had obtained using a further {\em ad hoc} approximation of treating the two noise terms as the same. The difference between equation (\ref{eqn:simplified}) and equation (\ref{eq:lowkappa2}) is proportional to $2\omega^2q_{\parallel}^2 (2\mathcal{E}^2-q^2)+2 D^4 q_{\parallel}^2\big[\mathcal{E}^4 (q_{\perp}^2-q_{\parallel}^2)-q^2 (q^4+2\mathcal{E}^2q_{\parallel}^2)\big]$, which is a collection of terms with positive and negative contributions that happen to lead to a small net quantitative contribution in the range of parameters that are relevant to our system. Therefore, equation (\ref{eqn:simplified}) might be used as a simplified compact expression that captures the essence of the phenomenon, although it is not an exact description.

\section{Concluding Remarks}
A comprehensive quantitative description of the linearized theory of ionic current fluctuations is presented and verified using Brownian dynamics simulations. For typical conditions, the derived expression for the spectrum of current fluctuations features a plateau at low frequencies and a power law decay at higher frequencies before crossing over to the white noise at very high frequencies. The effect of various control parameters and the resulting changes in the behaviour of the current fluctuations was studied through an analysis of the asymptotic limiting forms of the obtained expression. We expect our work to be useful for characterizing the statistics of current fluctuations in nano-pores. 

\section*{Acknowledgment}
We would like to thank Douwe Bonthuis who was involved during the early stages of this work.

\section*{References}

\providecommand{\newblock}{}

\end{document}